\documentclass[a4paper,12pt]{article}
\usepackage{t1enc}
\usepackage{times,euler}
\usepackage{amsmath,amsfonts,amssymb}
\usepackage[latin2]{inputenc}
\usepackage{graphicx}

\newcommand{\be}{\begin{eqnarray}}
\newcommand{\ee}{\end{eqnarray}}

\newcommand{\bi}{\begin{itemize}}
\newcommand{\ei}{\end{itemize}}
\newcommand{\bc}{\begin{center}}
\newcommand{\ec}{\end{center}}
\newcommand{\ud}{\mathrm{d}}

\begin{document}

{\bf{\Large
Thermodynamic interpretation of the uniformity of the phase space probability measure}}

\vspace{7mm}

{\bf W. Wi\'slicki}

\vspace{5mm}
\noindent
A. So\l tan Institute for Nuclear Studies, Ho\.za 69, PL-00-681 Warsaw \\
\noindent
and \\
\noindent
Interdisciplinary Centre for Mathematical and Computational Modelling, \\
University of Warsaw, Pawi\'nskiego 5a, PL-02-160 Warsaw, \\

\noindent
email: wislicki@fuw.edu.pl

\begin{abstract}
Uniformity of the probability measure of phase space is considered in the framework of classical equilibrium thermodynamics.
For the canonical and the grand canonical ensembles, relations are given between the phase space uniformities and thermodynamic potentials, their fluctuations and correlations. 
For the binary system in the vicinity of the critical point the uniformity is interpreted in terms of temperature dependent rates of phases of well defined uniformities.
Examples of a liquid-gas system and the mass spectrum of nuclear fragments are presented.
\end{abstract}

\vspace{5mm}


\vspace{5mm}

{\bf {\large Introduction}}

\vspace{3mm}

Generalized entropies were proposed by Renyi \cite{renyi70} to characterize probability measures.
Given a probability measure $\ud\mu$ and a partitioning of a phase space into $M(\Delta)$ cells, $\{\Delta_i\}_{i=1}^{M(\Delta)}$, each of the same volume $\Delta$, the Renyi entropies $I_q$ ($q\in{\boldmath R^1}$) are defined
\begin{eqnarray}
I_q & = & \left\{\begin{array}{lc}
    \frac{1}{1-q}\ln \sum_{i=1}^{M(\Delta)} p_i^q, & \;\;\;\;\; q\neq 1 \\
    & \\
    -\sum_{i=1}^{M(\Delta)} p_i\ln p_i, & \;\;\; q=1
           \end{array}\right . 
\label{one}
\end{eqnarray}
where
\begin{eqnarray}
p_i & = & \int_{\Delta_i} \ud\mu 
\end{eqnarray}
and sums run over cells for which $p_i\neq 0$.
It has been found later by many authors \cite{allofthem} that for investigation of singular and multifractal measures a useful quantity is a generalized dimension $D_q$, related to the Renyi entropy $I_q$
\begin{eqnarray}
D_q=-\lim_{\Delta\rightarrow\infty}\frac{I_q}{\ln\Delta}.
\end{eqnarray}
The Renyi dimension $D_q$ is always positive and decreases with $q$, so that there always exists $D_{\infty}=\lim_{q\rightarrow\infty}D_q$.
For positive $q$, the $D_q$ and $I_q$ is mostly sensitive to these regions of the phase space where the measure is concentrated and for negative $q$ -- where it is rarified.

In order to find a universal description of probabilty distributions for multifractals, another $q$-dependent quantity, called uniformity, was constructed by Beck in ref. \cite{beck90} from Renyi dimensions or entropies
\begin{eqnarray}
\gamma_q=\frac{D_q+q(q-1)D_q'}{D_q+(q-1)D_q'}
\label{three}
\end{eqnarray}
where $'$ stands for derivative over $q$.
For all $q$, $0\leq\gamma_q\leq 1$.
It can be found solving eq. (\ref{three}) that for the most uniform probabily density $\gamma_q=1$ and 
\begin{eqnarray}
D_q=D_{\infty} 
\end{eqnarray}
does not depend on $q$.
The minimum uniformity $\gamma_q=0$ corresponds to 
\begin{eqnarray}
D_q=\frac{1}{1-1/q}D_{\infty}.
\end{eqnarray}
The derivative of the uniformity
\begin{eqnarray}
\gamma_q'=(q-1)\frac{D_q[2D_q'-(1-q)D_q'']}{[D_q-(1-q)D_q']^2}
\label{three1}
\end{eqnarray}
vanish for either $q=1$, if it is differentiable there, or for
\begin{eqnarray}
D_q=D_{\infty}+\frac{D_0-D_{\infty}}{1-q}.
\end{eqnarray}

The uniformity was introduced in order to characterize random variables on events generated by formal systems.
It seems interesting to apply this concept to random variables related to physical systems in various conditions and represented by variables of thermodynamical ensembles.
Since the probability measures for such ensambles are well defined and they determine both Renyi dimensions and all traditional thermodynamic potentials, the relations between these quantities should be easy to find.
On this way one could expect to gain some understanding on how the structure of the phase space is reflected in such global characteristics of the system as its thermodynamic functions.
Moreover, because of sensitivity of $D_q$ and $\gamma_q$, controlled by the parameter $q$, to singularities of the measure, their relation to fluctuations of some measurable quantities, as internal energy or numbers of particles, or correlations of those, has to be straightforward.  
It can be also expected that two control parameters, the $q$ for Renyi entropies and the temperature for thermodynamic potentials, should to some extent play similar roles in the formalism, the latter determining the strength of thermal fluctuations and the former steering the sensitivity of $D_q$ and $\gamma_q$ to them. 

\vspace{7mm}

{\bf{\large The Canonical Ensemble}}

\vspace{3mm}

Consider the canonical ensemble with the probability measure 
\begin{eqnarray}
\ud\mu=\frac{1}{Z(\beta)}e^{-\beta E}\ud E
\end{eqnarray}
where $E$ stands for energy and 
\begin{eqnarray}
Z(\beta)=\int \ud\mu 
\end{eqnarray}
is the partition function, where the integral is taken over the whole phase space, $\beta=1/T$ is inverse temperature and the Boltzmann constant is assumed equal one.
Using (\ref{one}) we find for $\Delta\rightarrow 0$
\begin{eqnarray}
D_q\ln\Delta & = & \left\{\begin{array}{lc}
\frac{F(T/q)-F(T)}{T(1/q-1)}, & \;\;\;\;\; q\neq 1 \\
 & \\
\frac{\partial F(T)}{\partial T}=-S(T), & \;\;\;\;\; q=1
                   \end{array}\right .
\label{five}
\end{eqnarray}
where
\begin{eqnarray}
F(T)=-T\ln Z(T)
\end{eqnarray}
is free energy. 
Relation similar to $(\ref{five})$, between Renyi entropies and free energy, was found by T\'el in different context in ref. \cite{tel88}.
It is seen from eq. (\ref{five}) that $q$ is temperature scaling parameter.
For $q\neq 1$ the Renyi entropy $I_q$ is equal to the finite difference ratio of $F$ and $T$ and for $q\rightarrow 1$ it becomes the usual relation between $F$ and the thermodynamic entropy $S$.
Using (\ref{five}), the uniformity (\ref{three}) for the canonical ensemble is found
\begin{eqnarray}
\frac{1}{\gamma_q}=\frac{1}{q}-\frac{F(T)-F(T/q)}{TS(T/q)}.
\end{eqnarray}
The maximum uniformity $\gamma_q=1$ corresponds to
\begin{eqnarray}
qF(T/q) & = & F(T)=-TS_{\mbox{\scriptsize max}}
\end{eqnarray}
where $S_{\mbox{\scriptsize max}}$ is the maximum entropy.
For the minimum uniformity, $\gamma_q=0$, the free energy $F$ does not depend on $q$ and 
\begin{eqnarray}
F(T/q) & = & F(T)=-TS_{\mbox{\scriptsize min}},\;\;\;\;\;q>1
\label{seven1}
\end{eqnarray}
where $S_{\mbox{\scriptsize min}}$ is the minimum entropy.

Calculating the first and the second derivatives of $D_q$ and substituting to eq. $(\ref{three1})$ one finds
\begin{eqnarray}
\gamma_q'(T)=\frac{1}{T/q}\frac{F(T/q)-F(T)}{[U(T/q)-F(T)]^2}{\cal V}[E(T/q)],
\end{eqnarray}
where ${\cal V}(E)$ is the variance of the energy $E$ and $U=\langle E\rangle$ is internal energy, or rewriting it in terms of the specific heat $c_V(T/q)={\cal V}[E(T/q)]/(T/q)^2$, using eq. $(\ref{five})$ and formula $F=U-TS$
\begin{eqnarray}
\gamma_q'(T) & = & \left\{\begin{array}{lc}
0, & \;\;\;\;\; q > 1 \\
& \\
\frac{c_V(T/q)(1-q)S(T)}{[(1-q)S(T)-S(T/q)]^2}, & \;\;\;\;\; q\rightarrow 1^-
                   \end{array}\right .
\end{eqnarray}

\vspace{7mm}

{\bf{\large The Grand Canonical Ensemble}}

\vspace{3mm}

For the grand canonical ensemble the probability measure depends in addition on the number of particles $N$
\begin{eqnarray}
\ud\mu_N=\frac{1}{\Xi(\beta,\mu_c)}e^{\beta[\mu_cN-E_N]}\ud E_N
\end{eqnarray}
where $\mu_c$ is the chemical potential and $\Xi$ is the grand canonical sum
\begin{eqnarray}
\Xi(\beta,\mu_c)=\sum_{N=0}^{\infty}\int \ud\mu_N.
\end{eqnarray}
The integral is taken over the whole $N$-particle phase space.

The Renyi dimensions in this case are ($\Delta\rightarrow 0$)
\begin{eqnarray}
D_q\ln\Delta & = & \left\{\begin{array}{lc}
\mu_c\frac{\langle N(T/q)\rangle-\langle N(T)\rangle}{T(1/q-1)}, & \;\;\;\;\; q\neq 1 \\
 & \\
-\mu_c\frac{\partial \langle N(T)\rangle}{\partial T}=-S(T), & \;\;\;\;\; q=1
                   \end{array}\right .
\label{nine}
\end{eqnarray}
where the average number of particles is given by
\begin{eqnarray}
\langle N(T)\rangle=T\frac{\partial \ln \Xi(T,\mu_c)}{\partial \mu_c}
\end{eqnarray}
and the uniformity is equal to
\begin{eqnarray}
\frac{1}{\gamma_q}=\frac{1}{q}+\mu_c\frac{\langle N(T)\rangle -\langle N(T/q)\rangle}{TS(T/q)}.
\end{eqnarray}
Analogously to the canonical ensamble, the condition for the maximum uniformity is
\begin{eqnarray}
q\langle N(T/q)\rangle & = & \langle N(T)\rangle=\frac{T}{\mu_c}S_{\mbox{\scriptsize max}}
\end{eqnarray}
and for the minimum uniformity is
\begin{eqnarray}
\langle N(T/q)\rangle & = & \langle N(T)\rangle=\frac{T}{\mu_c}S_{\mbox{\scriptsize min}},\;\;\;\;\; q>1.
\end{eqnarray}

Calculating the first and the second derivatives of $D_q$ one finds the derivative of $\gamma_q$
\begin{eqnarray}
\gamma_q'(T) & = & \frac{1}{T/q}\frac{\langle N(T/q)\rangle-\langle N(T)\rangle}{[U(T/q)-\mu_c(\langle N(T/q)\rangle-\langle N(T)\rangle)]^2}\times \nonumber \\
& & \nonumber \\
&  & \{{\cal V}[\mu_c\langle N(T/q)\rangle]+{\cal V}[E(T/q)]-2\mu_c\mbox{cov}[\langle N(T/q)\rangle,E(T/q)]\}.
\end{eqnarray}
In case of constant volume and using eq. $(\ref{nine})$, the formula can be rewritten in the vicinity of $q=1$ in terms of the specific heat $c_V$ and isothermal compressibility $\kappa_T(T/q)={\cal V}[\langle N(T/q)\rangle]/p\langle N(T/q)\rangle$
\begin{eqnarray}
\gamma_q'(T) & = & \left\{\begin{array}{lc}
0, & \;\;\;\;\; q > 1 \\
& \\
\frac{(1-q/T)S(T)}{[T/qS(T/q)+\mu_c\langle N(T)\rangle]^2} \times \\
\{\mu_cp\kappa_T(T/q)\langle N(T/q)\rangle+(T/q)^2c_V(T/q)- \\
2\mu_c\mbox{cov}[\langle N(T/q)\rangle,E(T/q)]\}, & \;\;\;\;\; q\rightarrow 1^-
                   \end{array}\right .
\end{eqnarray}

\vspace{7mm}

{\bf{\large Applications and Discussion}}

\vspace{3mm}

The uniformity parameter $\gamma_q$ was originally proposed \cite{beck90} to measure distance from the two extremes: the chaotic phase ($\gamma_q=1$) and condensed phase ($\gamma_q=0$).
We show in the following that for the system consisting of subsystems in thermal equilibrium, the overall uniformity is the weighted average of uniformities of subsystems with weights given by particle numbers ratios for subsystems.

Consider the two--phase system with average numbers of particles $\langle N_{1(2)}\rangle$ and the uniformities of pure phases $\gamma_{1(2)}$.
In order to find the condition for $\gamma_{1(2)}$ to be independent of $q$ one solves eq. (\ref{three})
\begin{eqnarray}
\gamma_{1(2)}=\frac{D_{1(2)q}+q(q-1)D_{1(2)q}'}{D_{1(2)q}+(q-1)D_{1(2)q}'}
\end{eqnarray}
with respect to $D_q$ 
\begin{eqnarray}
D_{1(2)q} & = & D_{1(2)\infty}\frac{q-\gamma_{1(2)}}{q-1} \nonumber \\
D_{1(2)q}' & = & D_{1(2)\infty}\frac{\gamma_{1(2)}-1}{(q-1)^2}.
\label{thirteen}
\end{eqnarray}
Using eq. (\ref{nine}) and the fact that chemical potentials in subsystems are equal in equilibrium, one finds
\begin{eqnarray}
D_q=D_{1q}+D_{2q} 
\end{eqnarray}
and, using (\ref{thirteen}),
\begin{eqnarray}
\gamma_q(T,\mu_c)=c_1(T,\mu_c)\gamma_1+c_2(T,\mu_c)\gamma_2, 
\end{eqnarray}
where 
\begin{eqnarray}
c_{1(2)}=\frac{\langle N_{1(2)}\rangle}{\langle N_1\rangle +\langle N_2\rangle}
\end{eqnarray}
are concentrations of particles in subsystems 1 and 2.
Generalization to more than two subsystems is straightforward.

\vspace{3mm}

{\it {\large Example 1: classical liquid-gas system}}

Consider any classical system of two coexisting phases in thermal equilibrium near the critical point.
In the vicinity of critical temperature $T_c$, phase densities follow the power--law dependence on temperature \cite{toda}
\begin{eqnarray}
\rho_2-\rho_1=\mbox{const}\cdot (T_c-T)^{\alpha},\;\;\;\;\;T<T_c
\label{sixteen}
\end{eqnarray}
where $\alpha$ is the critical exponent (e.g. for the liquid-gas system its value is between 0.33 and 0.36).
Assuming the average total number of particles $\langle N_1+N_2\rangle$ independent of temperature and the volume of the system to be constant, from eq. (\ref{sixteen}) follows that
\begin{eqnarray}
c_1 & = & A+B\left(1-\frac{T}{T_c}\right)^{\alpha} \nonumber \\
c_2 & = & 1-c_1
\label{seventeen}
\end{eqnarray}
where $A$ and $B$ are constants.
As discussed before, the parameter $q$ plays the role of the temperature scaling parameter.
Thus the temperature evolution in eq. (\ref{seventeen}) can be rewritten in terms of $q=T/T_c$ \footnote{The choice of $T_c$ as the reference is arbitrary but convenient. For another temperature scale one has to change the boundary conditions in eqns. (\ref{seventeen-p})}.
Then $q=1$ corresponds to the critical temperature and domains $q>1$ and $0<q<1$ correspond to pure uniform (gas) and mixed phases, whereas the limit $q=0$ should represent pure condensed (liquid) phase.
For temperature $T$ much lower than $T_c$ there is only condensed phase in the system and a weak dependence of density on $T$ can be neglected.
Therefore, for $q_0=T/T_0 < 1$, one finds the values of $A$ and $B$ from the limits 
\begin{eqnarray}
\lim_{q\rightarrow 1^-}\gamma_q & = & \gamma_1 \nonumber \\
\lim_{q\rightarrow q_0}\gamma_q & = & \gamma_2 .
\label{seventeen-p}
\end{eqnarray}
Hence, from (\ref{seventeen})
\begin{eqnarray}
c_1 & = & 1-\frac{1}{1-\gamma_2/\gamma_1}\left(\frac{1-q}{1-q_0}\right)^{\alpha} \nonumber \\
& & \nonumber \\
c_2 & = & \frac{1}{1-\gamma_2/\gamma_1}\left(\frac{1-q}{1-q_0}\right)^{\alpha}
\label{eighteen}
\end{eqnarray}
for $q_0\leq q<1$ and 
\begin{eqnarray}
c_{1(2)} & = & 0 \nonumber \\
c_{2(1)} & = & 1
\end{eqnarray}
for $q<q_0$($\geq 1$).
The $\gamma_q$ does not have to be differentiable at $q=1$.

The uniformity $\gamma_q$ for $\alpha=0.34$ and $c_{1,2}$ given by (\ref{eighteen}) is shown in fig.~1. 
\begin{figure}[h]
\begin{center}
\includegraphics*[width=90mm,height=70mm]{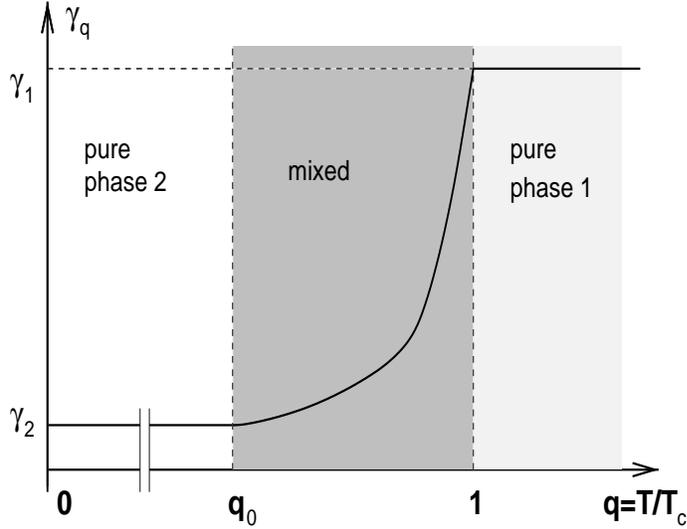}
\end{center}
\caption{\em The uniformity $\gamma_q$ for the two-phase system, assuming the near-to-critical power-law behaviour of densities difference as functions of temperature. Critical exponent $\alpha=0.34$ is assumed.}
\end{figure}
Since the power-law dependence of densities difference on temperature is valid for the near-to-critical region, this picture is realistic only for $T\rightarrow T_c$ or $q\rightarrow 1$ (see broken horizontal axis).
However, if temperature dependence of phase densities is known for any temperature, the same line of reasoning can be followed to find the overall uniformity for the mixture of phases in wide range of temperatures.
For some models, e.g. the power--law probability density mentioned in ref. \cite{beck90}, the uniformities of pure phases can be exactly calculated.
In other cases, provided the probability measure is known, these pure uniformities can be also determined by direct, although sometimes complex and time consuming, computation.
On the other hand, if the phase space probability density in the mixed--phase domain is measured experimentally and the uniformities of pure phases are known from elsewhere, or can be calculated, this theoretical scheme can be used to determine the ratio of phases at given temperature.

\vspace{3mm}

{\it {\large Example 2: discrete power-law energy spectrum}}

Let us discuss the system of discrete energy spectrum, where the energy $E$ of the system is a random variable taking positive integer values and exhibiting the power-law probability measure
\begin{eqnarray}
\ud\mu_E=\frac{E^{-\tau}}{\zeta(\tau)}\delta(E-m)\ud E, \;\;\;\;\;\;\; m=1,2,...
\end{eqnarray}
where $\zeta(\tau)=\sum_{m=1}^{\infty}m^{-\tau}$, $(\tau>1)$ is Riemann zeta function.
Expected value of the energy is then 
\begin{eqnarray}
\langle E\rangle =\frac{\zeta(\tau-1)}{\zeta(\tau)}
\label{nineteen}
\end{eqnarray}
and exists only for $\tau>2$.
Such statistical ensemble would be interesting {\it per se} and similar study of its properties in terms of Renyi entropy and Beck uniformity, as for the canonical and grand canonical ensembles, could be performed.  

In order to stay within the framework of equilibrium thermodynamics and use the concept of equilibrium temperature, consider the canonical case. 
The statistical sum is equal to
\begin{eqnarray}
Z(T) & = & \sum_{m=1}^{\infty}e^{-m/T} \nonumber \\
     & = & \frac{1}{e^{1/T}-1}.
\end{eqnarray}
Internal energy is equal to
\begin{eqnarray}
U(T) & = & -\frac{\partial\ln Z(T)}{\partial(1/T)} \nonumber \\
     & = & \frac{1}{1-e^{-1/T}}
\end{eqnarray}
and by requiring it to be equal to $\langle E\rangle$ $(\ref{nineteen})$ we find the equiliblium temperature
\begin{eqnarray}
T_{eq}=-\frac{1}{\ln(1-1/\langle E\rangle)}
\label{twenty}
\end{eqnarray}
and the equilibrium probability measure
\begin{eqnarray}
\ud\mu_E=\frac{(1-1/\langle E\rangle)^m}{\langle E\rangle -1}\delta(E-m)\ud E,\;\;\;\;\;\;\; m=1,2,....
\end{eqnarray}
For this probability density we find $(\Delta\rightarrow 0)$
\begin{eqnarray}
D_q(T_{eq})\ln\Delta & = & \left\{\begin{array}{lc}
-\frac{1}{1-q}\ln[\langle E\rangle ^q-(\langle E\rangle -1)^q], & \;\;\;\;\; q\neq 1 \\
 & \\
(1-\langle E\rangle)\ln(\langle E\rangle-1)+\langle E\rangle\ln\langle M\rangle, & \;\;\;\;\; q=1
                   \end{array}\right .
\end{eqnarray}
and
\begin{eqnarray}
\gamma_q(T_{eq}) & = & \left\{\begin{array}{lc}
q-\frac{[\langle E\rangle ^q-(\langle E\rangle -1)^q]\ln[\langle E\rangle ^q-(\langle E\rangle -1)^q]}{\langle E\rangle ^q\ln \langle E\rangle-(\langle E\rangle -1)^q\ln (\langle E\rangle -1)}, & \;\;\;\;\; q\ne 1 \\
& \\
1, & \;\;\;\;\; q=1
                       \end{array}\right .
\label{twentyone}
\end{eqnarray}

\begin{table}[h]
\caption{\em Power-law exponents, temperatures, mean energies and uniformities for nuclear fragmentation data.}
\begin{center}
\begin{tabular}{|c|c|c|c|c|} \hline
$\tau$ & $\langle E\rangle $ & $T (MeV)$ & $q=T/T_c$ & $\gamma_q$ \\ \hline 
4.1 & 1.10 & 6.0 & 0.50 & $0.29\pm 0.05$ \\
3.8 & 1.14 & 6.2 & 0.52 & $0.36\pm 0.07$ \\
3.7 & 1.15 & 6.7 & 0.56 & $0.49\pm 0.08$ \\
3.0 & 1.37 & 7.2 & 0.60 & $0.65\pm 0.11$ \\
2.8 & 1.51 & 7.7 & 0.64 & $0.75\pm 0.12$ \\
3.2 & 1.28 & 8.2 & 0.68 & $0.79\pm 0.12$ \\
2.6 & 1.74 & 8.3 & 0.69 & $0.83\pm 0.13$ \\
2.4 & 2.21 & 14.0 & 1.16 & $0.98\pm 0.12$ \\
2.9 & 1.43 & 14.5 & 1.20 & $0.96\pm 0.11$ \\
2.9 & 1.43 & 15.0 & 1.25 & $0.95\pm 0.11$ \\ \hline
\end{tabular}
\end{center}
\end{table}
\begin{figure}[h]
\begin{center}
\includegraphics*[width=90mm,height=70mm]{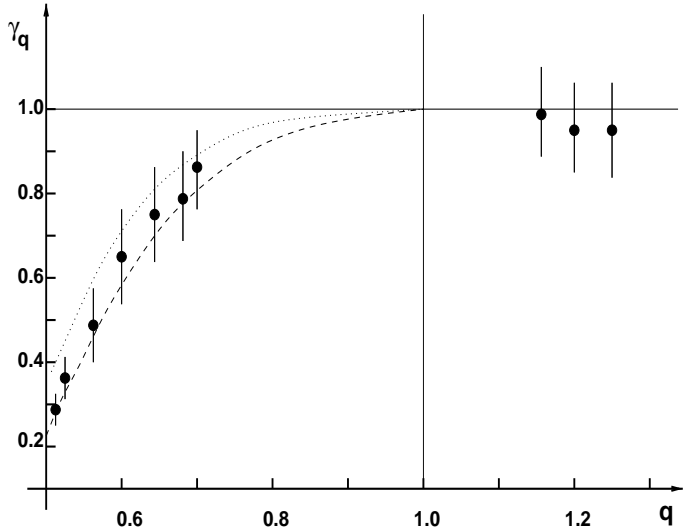}
\end{center}
\caption{\em The uniformity $\gamma_q$ for experimental data from nuclear fragmentation. The dashed line represents analytical uniformity for $\tau=4.1$ and the dotted one for $\tau=2.1$.}
\end{figure}
As an example of a real system exhibiting the power-law energy spectrum we employ the nuclear mass spectrum of heavy nuclei fragmentation.
Taking simplifying assumptions that the whole energy is identical to the fragment mass, neglecting the surface energy, the binding energy and finite size effects and assuming that all states are distinguishable, we use the power-law exponents determined experimentaly for low mass fragments and refered to in ref. \cite{frag}. 
The power-law behaviour is also characteristic for liquid cluster size distribution near the liquid-gas transition point in real systems and in percolation models. 
It has to be noted however, that $\tau$ exponent obtained for nuclear systems disassembly \cite{frag} is not a universal critical exponent but an apparent exponent absorbing some temperature dependence, usually unknown.
Nuclear data were analysed in ref. \cite{frag} in the framework of the condensation theory and the critical temperature $T_c=12.0\pm 0.2$ MeV was found.
For each data sample exhibiting the power-law mass distribution the temperature was determined from either the slope of the energy spectrum at $90^{\circ}$ or from the Fermi gas model or from the moving source model.
We identify this temperature with the equilibrium temperature $(\ref{twenty})$, define $q=T/T_c$ and calculate $\gamma_q(T)$ from eq. $(\ref{twentyone})$.
The numbers are given in tab.~1 and the uniformity as a function of $q$ in fig.~2.
For the calculation of errors of $\gamma_q$ the values for errors of $T_c$, $T$ and $\tau$ are taken from ref. \cite{frag}.
The curves in fig.~2 represent uniformities calculated using analytical formula $(\ref{twentyone})$ for $\tau=2.4$ and $\tau=4.1$.
As all experimental points stay within the belt determined by these curves we conclude that the temperature evolution given by $(\ref{twentyone})$ is consistent with data.
The data show that uniformities increase with temperature in the mixed-phase region and are consistent with 1 above critical temperature.
Below $T=4$ MeV the contribution of the non-uniform phase is below $10\%$.


\vspace{5mm}

\begin{thebibliography}{99}
\bibitem{renyi70} A. Renyi, {\it Probability Theory}, (Akademiai Kiado, Budapest, 1970)
\bibitem{allofthem} B.B. Mandelbrot, J. Fluid Mech. {\bf 62}, 331-342 (1974);\\ H.G.F. Hentschel and I. Procaccia, Physica {\bf D8}, 435-444 (1983);\\ P. Grassberger, Phys. Lett {\bf A97}, 227-230 (1983);\\ T.C. Halsey et al., Phys. Rev. {\bf A33}, 1141-1151 (1986)
\bibitem{beck90} C. Beck, Physica {\bf D41}, 67-78 (1990)
\bibitem{tel88} T. T\'el. Z. Naturforsch. {\bf 43a}, 1154-1174 (1988)
\bibitem{toda} M.~Toda, R.~Kubo and N.~Sait$\hat o$, {\it Statistical Physics I, Equilibrium Statistical Mechanics}, (Springer-Verlag, Berlin, Heidelberg, New York, Tokyo, 1983) 
\bibitem{frag} A.G. Panagiotou et. al., Phys. Rev. Lett. {\bf 52}, 496-500 (1984);
\end{thebibliography}
\end{document}